\documentclass[twocolumn,showpacs,floatfix,prl]{revtex4}

\usepackage{float}
\usepackage{graphicx}

\begin{document}

\title{Electron Teleportation via Majorana Bound States in a Mesoscopic Superconductor}

\author{Liang Fu}
\affiliation{Department of Physics, Harvard University, Cambridge, MA 02138}

\begin{abstract}
Zero-energy Majorana bound states in superconductors have been proposed to be 
potential building blocks of a topological quantum computer, because quantum information can be encoded in the fermion occupation of
a pair of {\it spatially separated} Majorana bound states.  
However, despite intensive theoretical efforts, non-local signatures of Majorana bound states have not been found in charge transport. 
In this work, we predict a striking non-local phase-coherent electron transfer process by virtue of tunneling in and out of a pair of 
Majorana bound states. This teleportation phenomena only exists in a mesoscopic superconductor because of an all-important but previously over-looked 
charging energy. We propose an experimental setup to detect this phenomena in a superconductor/quantum spin Hall insulator/magnetic insulator hybrid system. 
\end{abstract}

\maketitle

Majorana bound states are localized zero-energy excitations of a superconductor\cite{read, jackiw}. 
An isolated Majorana bound state is an equal superposition of electron and hole excitations and therefore
{\it not} a fermionic state. Instead {\it two} spatially seperated Majorana bound states together make {\it one} zero-energy fermion level\cite{read, ivanov} 
which can be either occupied or empty. This defines a two-level system which can store quantum information non-locally. 
 as needed to realize topological quantum computation\cite{kitaev, review}. 
While several schemes have been recently proposed to detect the {\it existence} of individual Majorana bound states\cite{demler, sarma, simon, bk1, fk1, lee, sau}, 
experimental signatures of the non-local {\it fermion occupation} of these states remain to be found. 

In this work, we predict a non-local electron transfer process due to Majorana bound states in a mesosopic superconductor: 
an electron which is injected into one Majorana bound state can go out from another one far apart 
{\it maintaining phase coherence}. Strikingly, the transmission phase shift is {\it independent} of the distance ``travelled''. In such a sense, we call
this phenomena ``electron teleportation''. It occurs because of the non-local fermion occupation of Majorana bound states and the finite charging energy of a mesoscopic superconductor. The all-important role of charging energy in the study of Majorana fermions has not been recognized before.  
We propose a realistic scheme to 
detect the teleportation phenomena in a superconductor/quantum spin Hall insulator/magnetic insulator hybrid system, which 
have been recently shown to host Majorana bound states\cite{fk2,fk3}. 

\begin{figure}
\centering
\includegraphics[width=3in]{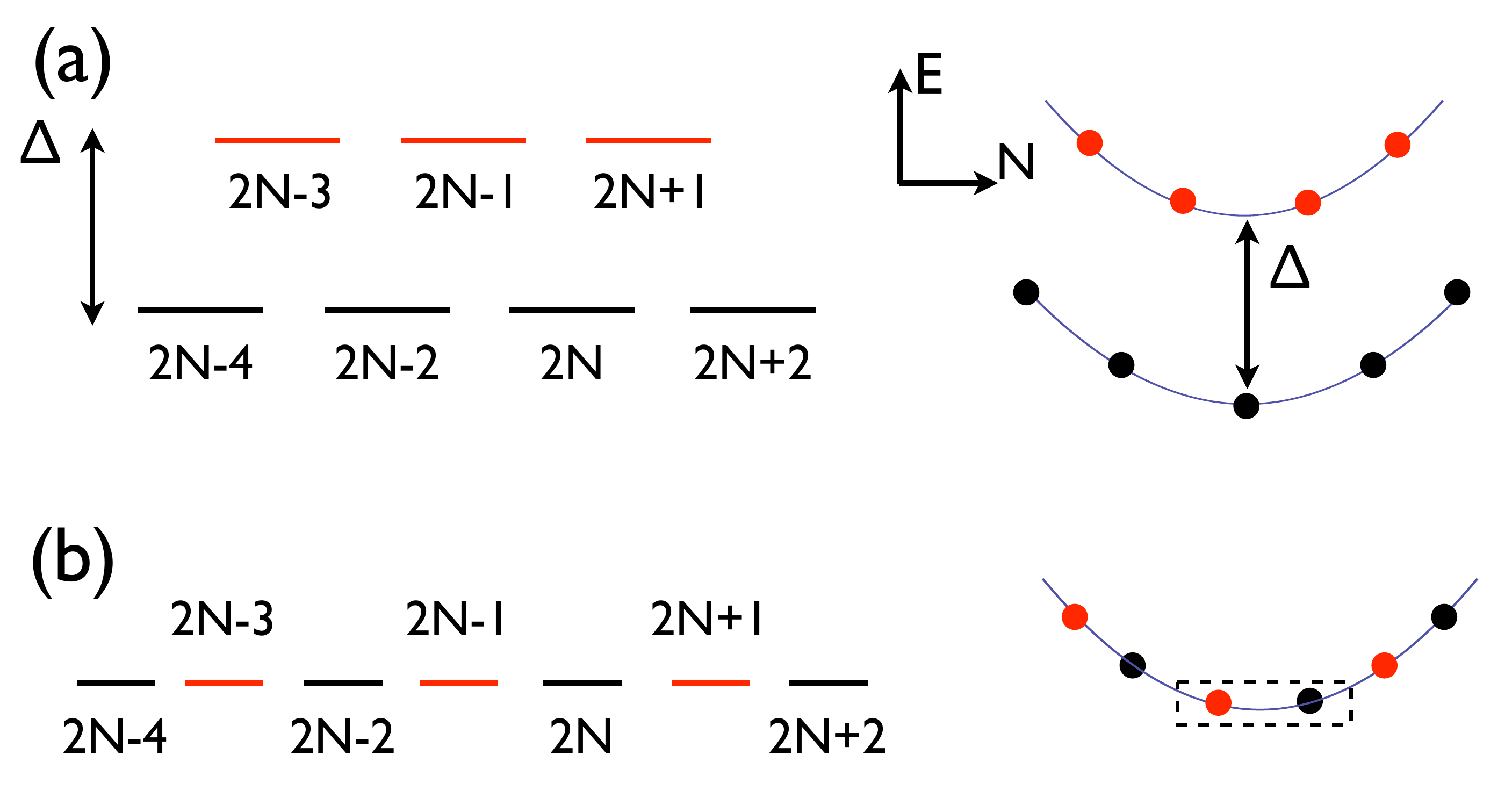}
\caption{Energy spectrum of a superconductor as a function of total number of electrons. States with an even and odd number of 
electrons are marked in black and red respectively. 
(a) and (b) correspond to superconductors without and with a pair of zero-energy Majorana bound states.
Figures on the left and right correspond to superconductors without and with charging energy. 
}
\end{figure}

In a macroscopic $s$-wave superconductor, charge $e$ excitations have a pairing energy gap, whereas 
charge $2e$ excitations cost zero energy. 
Therefore the ground state manifold consists of states with an even number of electrons only, as shown in Fig.1a.  
The BCS wave-function of the ground state with a definite {\it overall} superconducting phase $\phi \in [0, 2\pi]$, 
is a linear superposition of states with $2N$ electrons. 
Now consider that a pair of zero-energy Majorana bound states are present at positions $R_1$ and $R_2$ in the superconductor, 
and all other quasi-particle excitations have a finite gap greater than an energy scale $\Delta$. We shall show later how this situation can be realized in a device consisting of an $s$-wave superconductor and the recently discovered quantum spin Hall (QSH) insulator HgTe quantum well\cite{hgte,bhz}. 
The two Majorana operators $\gamma_1$ and $\gamma_2$ are defined by:
\begin{equation}
\gamma_{1,2} \equiv \int dx e^{-i\phi/2 }\xi_{1,2}(x) c^{\dagger}(x) + e^{i\phi/2 }\xi_{1,2}^*(x) c(x).
\label{mf}
\end{equation} 
Here $\xi_{1,2}(x)$ are bound state wave-functions centered at $R_{1,2}$. 
We assume that the distance between the two Majorana bound states is much larger than the coherence length---a necessary condition for the notion of non-locality to be meaningful.   
A single fermionic operator can then be defined $d^\dagger \equiv (\gamma_1 + i \gamma_2)/2$, which accommodates an extra fermion excitation 
without any energy cost. Now the ground states of the superconductor have two sectors $| e \rangle$ and $| o \rangle$ defined by $d | e \rangle =0, | o \rangle = d^\dagger | e \rangle$, which have an even and odd number of electrons respectively as shown in Fig.1b,  
\begin{eqnarray}
|e,\phi\rangle &=& \sum_{n=2N} a_{n} e^{i\phi N}  | 2N\rangle \nonumber \\
|o,\phi\rangle &=& \sum_{n=2N+1} a_{n} e^{i\phi (N+1/2)} | 2N+1 \rangle, \label{def}
\end{eqnarray}
where $a_n$ is real and slowly varying at large $n$. 
Eq.(\ref{def}) says that the fermion occupation of the $d$ level (empty or occupied) $d^\dagger d = (i \gamma_1 \gamma_2 +1)/2$, is {\it fixed} by the total number of electrons in the superconductor mod $2$: 
\begin{equation}
i\gamma_1 \gamma_2 = (-1)^n.  \label{gauge}
\end{equation}
Eq.(\ref{gauge}) imposes constraint on the Hilbert space. Equivalently, (\ref{gauge}) implements a gauge transformation $\gamma_j \rightarrow -\gamma_j, \phi \rightarrow \phi+2\pi$, which is a gauge symmetry in the definition of Majorana operators Eq.(\ref{mf}). 

If the superconductor under consideration is of mesoscopic size and connected to ground by a capacitor,
the energy spectrum has an additional term due to the finite charging energy:
\begin{equation}
U_c(n)=(ne-Q_0)^2/2C, \;\;n=0,\pm 1, \pm 2, ... \label{Uc}
\end{equation} 
where $Q_0$ is the gate charge determined by the gate voltage $V_g$ across the capacitor. 
As a result, states with different $n$ are no longer degenerate.
In this work we will consider the regime $U \equiv e^2/C  < \Delta $, which can always be satisfied by increasing the size of the superconductor. 
The low-energy spectrum ($E<U$) then depends {\it crucially} on whether Majorana bound states are absent or present. 
In the former, only states with an even number of electrons are important at low energy, which leads to an even-odd effect in tunneling experiments on mesoscopic superconductors\cite{tinkham}. 
In contrast, if Majorana bound states are present, both even- and odd-states appear on equal footing in the low-energy spectrum. 
In this case,  when $Q_0/e$ is adjusted to half-integers, an energy-level degeneracy can be achieved between two states that differ by charge $e$, instead of $2e$ as in a usual superconductor with Coulomb blockade. This two-level system shown in Fig.1 is the main subject of our study.

\begin{figure}
\centering
\includegraphics[width=3in]{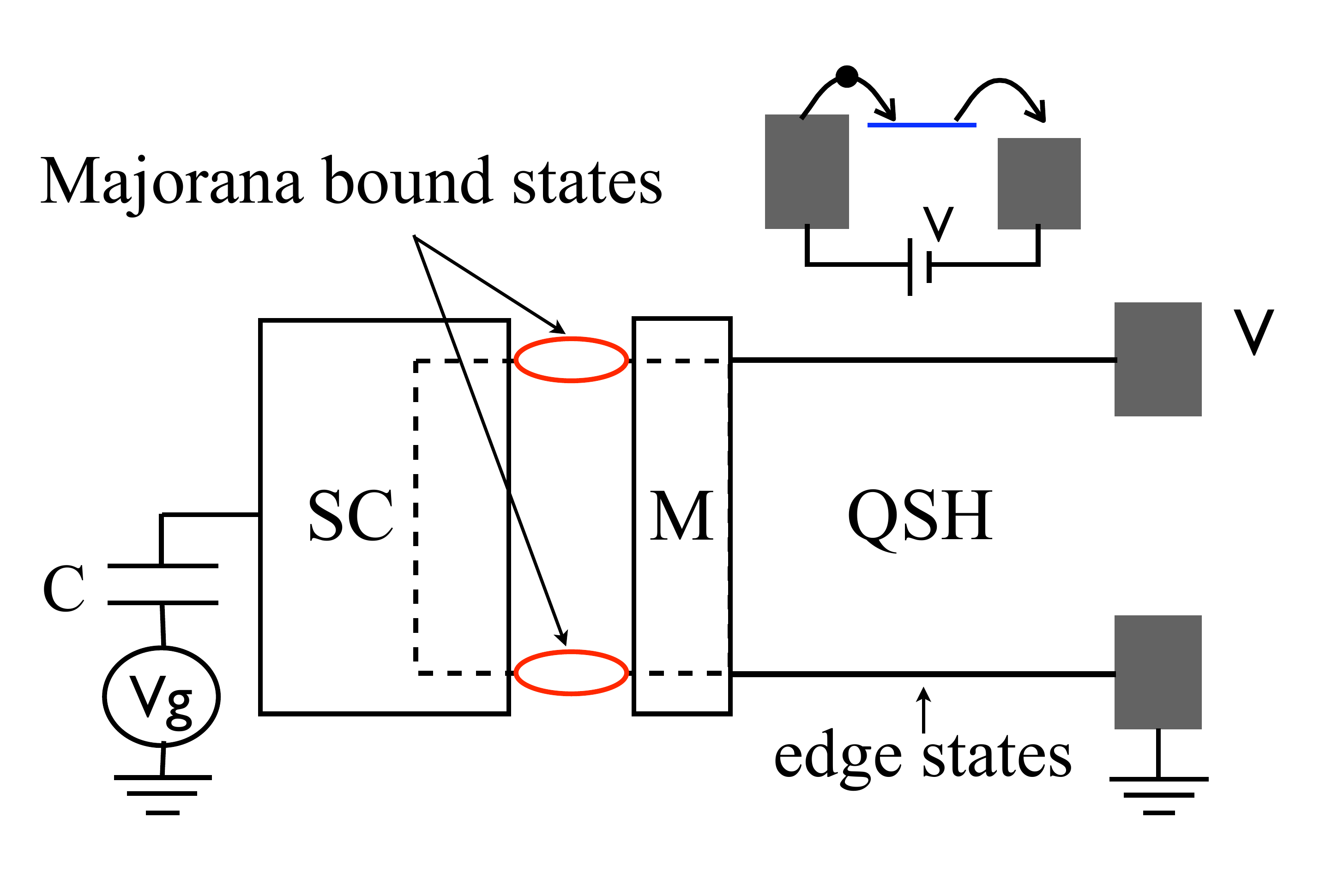}
\caption{the device used to study electron tunneling from leads into two Majorana bound states, consisting of an $s$-wave superconductor (SC) and a magnetic insulator (M) junction deposited on top of a quantum spin Hall insulator (QSH). 
By tuning the gate voltage across the capacitor, the superconductor is close to a charge degeneracy point between a total number of electron $n_0 $ and $(n_0+1)$.  This two-level system corresponds to a resonant level empty or occupied.  At a small bias, electron tunneling through {\it two} Majorana bound states is equivalent to {\it phase-coherent} tunneling through a {\it single} resonant level.  
}
\end{figure}

We now weakly couple the two Majorana bound states to separate normal metal leads. 
This can be realized in an $s$-wave superconductor/quantum spin Hall insulator/magnetic insulator hybrid system. 
The QSH insulator used here is a new phase of two-dimensional insulators which have robust edge states\cite{km}. It has been experimentally realized in HgTe quantum wells\cite{hgte}. The device geometry is shown in Fig.2: an $s$-wave superconductor and a magnet are deposited on top of the QSH insulator. 
Both proximity effect with the superconductor and the Zeeman field due to the magnet open up a finite quasi-particle gap for the QSH edge states. 
However, two zero-energy Majorana bound states $\gamma_{1,2}$ exist at the intersection of the superconductor-magnet interface with the top and bottom edge respectively\cite{fk3}, conceptually similar to the states localized at the ends of a one-dimensional spinless p-wave superconductor\cite{kitaev2}. 
The edge states in the uncovered part of the quantum spin Hall insulator are naturally used as leads to connect $\gamma_{1}$ and $\gamma_{2}$ 
to source and drain.    


To describe electron tunneling between the lead and the superconductor, we write the electron operator in terms of quasi-particle operators of the superconductor
\begin{eqnarray}
c(x) &=& e^{-i\phi/2} \left[ \xi_1(x) \gamma_1 +  \xi_2(x) \gamma_2 + ... \right].  
\label{electron}
\end{eqnarray}   
Since we will only consider small bias voltage $V<U<\Delta$, only zero-energy Majorana operators are important and contributions from other quasi-particle states can be neglected in (\ref{electron}). 
We now write down the Hamiltonian for the system in Fig.2: $H=H_L  + U_c 
+ H_T $, where $H_L=\sum_{k,j=1,2} \epsilon_j(k) c^\dagger_{j,k} c_{j,k}$ describes the two leads, $U_c$ is the charging energy defined in (\ref{Uc}). 
The effective tunneling Hamiltonian $H_T$ at low energy is given by substituting (\ref{electron}) into the bare tunneling term $t_i c_i^\dagger c$:
\begin{eqnarray}
H_T&=& \sum_{j=1,2}   \left[  \lambda_j c_j^\dagger   \gamma_j e^{-i\phi/2}  + \lambda_j^*  \gamma_j c_j e^{i\phi/2} \right]. 
\label{Ht}
\end{eqnarray}
Here $c_j$ annihilates an electron in lead $j$ and $\lambda_j \propto \xi_j(R_j)$ is the tunneling matrix element. As we emphasized earlier, $\xi_1(x)$ and $\xi_2(x)$ have essentially {\it zero} wavefunction overlap so that no coupling between $c_1$($c_2$) and  $\gamma_2$($\gamma_1$)  exists. The operator $e^{\pm i\phi/2}$ in $H_T$ increases/decrases the total charge of the  superconductor by one charge unit 
$[n, e^{\pm i\phi/2}] = \pm e^{\pm i \phi/2}$, 
and the Majorana operator $\gamma_{1,2}$ changes the parity of electron number in the superconductor. The ``naive'' Hilbert space of $H$ is simply the direct product of electron number eigenstate $|n\rangle$ and the two states of $d$-level ($|e\rangle$ and $|o\rangle$), but it is {\it redundant}.
Instead the physical Hilbert space only consists of those states $| n =2N; e \rangle $ and $ | n=2N+1; o \rangle$ that satisfies the gauge constraint (\ref{gauge}).

When the source is biased at a small voltage $V$, current flows to drain by electron tunneling in and out of the superconductor. Since charging energy $U_c$ favors states with a fixed number of charge in the superconductor,  
only two charge states $| n_0 \rangle$ and $| n_0 + 1 \rangle$ give dominant contribution to the current for $V < U$, which is similar to tunneling through a quantum dot in the Coulomb blockade regime. To a good approximation, we can then {\it truncate} the Hilbert space only keeping these two states, which we label by $s_z=\pm 1$. $H$ then becomes\begin{equation}
\tilde{H}= H_L + \frac{\delta}{2} s_z + \sum_{j=1,2}   \left[  \lambda_j c_j^\dagger   \gamma_j s_- +  \lambda_j^* \gamma_j c_j  s_+ \right].
\label{spin}
\end{equation}
Here $\delta$ is the energy difference between $| n_0 \rangle$ and $| n_0 + 1 \rangle$ and depends on the gate voltage. The gauge symmetry (\ref{gauge}) then becomes 
$i\gamma_1 \gamma_2 s_z=(-1)^{n_0}$.
The key to solve the tunneling problem (\ref{spin}) is to combine Majorana and spin operators into a {\it singe} fermion operator $f$.
\begin{eqnarray}
 \gamma_1 s^+  \rightarrow  f^+,&& \;  \gamma_1 s^-  \rightarrow  f \nonumber \\
\gamma_2 s^+  \rightarrow   i  (-1)^{n_0} f^+,&& \; \gamma_2 s^-  \rightarrow  i (-1)^{n_0+1} f.
\label{transform}
\end{eqnarray}
We have checked that this transformation preserves all commutation relations
\begin{eqnarray}
\{ \gamma_j s^+, \gamma_j s^- \}&=& 1, \nonumber \\ 
\{ \gamma_i s^+, \gamma_j s^+ \}&=&\{ \gamma_i s^-, \gamma_j s^- \}=0, \nonumber \\
\{ \gamma_1 s^+, \gamma_2 s^- \}&=& \gamma_1 \gamma_2 s_z = i(-1)^{n_0+1},
\end{eqnarray}
where the gauge constraint is used in the last equation. 
Conceptually, it is not surprising that the transformation (\ref{transform}) works: after all, the two charge states $| n_0 \rangle$ and $| n_0 + 1 \rangle$ differ by one electron.     
Using (\ref{transform}), we rewrite the Hamiltonian $\tilde{H}$ in terms of the fermion operator $f$:
\begin{eqnarray}
\tilde{H}= &&H_L+  \delta(f^\dagger f - \frac{1}{2})+ ( \lambda_1  c_1^\dagger  f + hc) \nonumber \\
&&+   (-1)^{n_0} (-i \lambda_2 c_2^\dagger  f + hc ). \label{resonant}
\end{eqnarray} 
Eqs.(\ref{Ht}, \ref{spin}, \ref{transform}, \ref{resonant}) are the main results of this work, and to the best of our knowledge they have not been reported before.
(\ref{resonant}) says that electron tunneling in and out of two {\it spatially separated} Majorana bound states is equivalent to resonant tunneling through a {\it single} level, as shown schematically in Fig.2. Since resonant tunneling is a coherent process, we conclude that an incident electron at $E<U$ tunnels into one Majorana bound state and comes out from its partner far apart still maintaining phase coherence. Strikingly, the magnitude and phase of the transmission amplitude---which we call $t$---is {\it independent} of their distance. In this sense, we call such a non-local electron transfer process ``teleportation''. 

The phase coherence over a  long distance shown here is in fact a direct consequence of Majorana bound states. Conceptually it can be best understood from electron's Green function
$G^{e(o)}(x,t; y,0) \equiv \langle  c(x,t) c^\dagger(y,0)   \rangle_{e(o)}$ defined in the even and odd ground state sector $|e\rangle$ and $|o\rangle$ respectively\cite{gauge}. 
Using (\ref{electron}), we find
\begin{eqnarray}
G^{e,o}(x,t\rightarrow \infty; y,0)  = \mp i \xi_2^*(y) \xi_1(x)   \sim  O(1) \label{Green}
\end{eqnarray}  
is finite for $x \sim R_1, y \sim R_2$. The long-time limit corresponds to the low-bias regime we are interested in. The fact that (\ref{Green}) is independent of $|R_1-R_2|$ is most unusual, as first pointed out in Ref.\cite{semenoff}.

\begin{figure}
\centering
\includegraphics[width=3.5in]{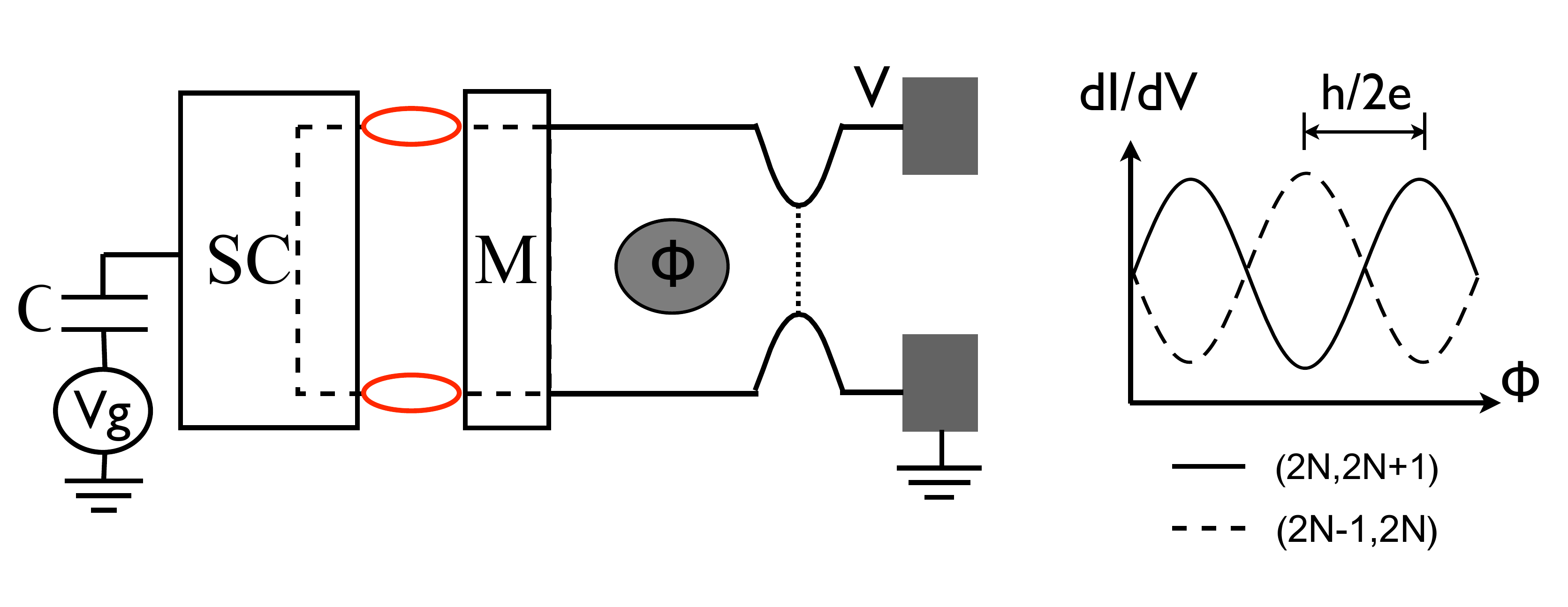}
\caption{Left: an interferometer that probes the {\it phase-coherent} electron teleportation via two Majorana bound states. 
Right: schematic plot of zero bias differential conductance as a function of magnetic flux at two successive charge degeneracy points $(2N-1, 2N)$ 
and $(2N, 2N+1)$. The $h/2e$ shift in conductance peak signals the change of fermion number parity in the superconductor.}
\end{figure}

We now show that, interestingly, the phase of transmission amplitude $t$ depends on the gate charge $Q_0$ in a surprising way, 
and therefore it is sensitive to the fermion occupation ($|e\rangle$ vs. $|o\rangle$) of the Majorana bound state pair. Consider tuning gate voltage to make $Q_0$ change by one charge unit. The number of electrons in the  superconductor ground state will correspondingly change by one. Although the excitation energy spectrum $U_c$ comes back to itself, we find  
\begin{equation}
t \rightarrow -t, \textrm{ when } Q_0 \rightarrow Q_0 \pm e, \label{phase}
\end{equation} 
i.e., the transmission phase shift changes by $\pi$. This behavior is related to the change of fermion number parity in the ground state. The property (\ref{phase}) is evident from the $(-1)^{n_0}$ factor in $\tilde{H}$, which valid in the two-level approximation. Using second-order perturbation theory in the weak tunneling limit, one can easily show that this $(-1)^{n_0}$ factor comes from the $\pm$ sign in Eq.(\ref{Green}). In general, we can prove (\ref{phase}) using the following symmetry of the full Hamiltonian $H$ and the gauge constraint (\ref{gauge})
\begin{eqnarray}
U H(Q_0, \lambda_2) U^{-1} &=& H(Q_0 + e, -\lambda_2), \nonumber \\
U  i \gamma_1 \gamma_2 (-1)^n U^{-1} &=& i \gamma_1 \gamma_2 (-1)^n, \; U \equiv \gamma_2 e^{-i \phi/2}. \nonumber 
\end{eqnarray}




To detect the phase-coherence of the electron teleportation described above, 
we consider the interferometer setup in Fig.3: a point contact is introduced to partially scatter an incident electron at the top edge directly to the bottom edge, and partially transmit it to the superconductor which subsequently comes out at the bottom edge. Interference between the two paths with a magnetic flux $\Phi$ enclosed leads to a $\Phi$-dependent differential conductance $dI/dV$, which is $h/e$-periodic. If the direct scattering probability is large,  interference visibility is maximum at  the charge degeneracy point when electron tunneling through Majorana bound states is {\it on resonance}. 
We schematically plot the $\Phi$ dependence of $dI/dV$ (at zero bias and zero temperature) 
for two successive half-integer charges $Q_0$ in Fig.3. The sign reversal of $t$ discussed in (\ref{phase}) leads to a $h/2e$ shift in the interference pattern.


{\it Discussion}: It is instructive to compare our study of tunneling into Majorana bound states in the $V < U$ regime with previous studies 
which do not include the charging energy $U$\cite{demler, sarma, bk1}. Instead of using a ``floating'' superconductor as in Fig.2, 
these works consider a {\it grounded} superconductor in tunneling contact with
two leads each having an independent bias voltage. Such a {\it three-terminal} device can be realized in a geometry shown in Fig.4. 
In this setup, transferring two electrons to the superconducting condensate does not cost energy. Therefore, when the two Majorana bound states are sufficiently far apart, an incident electron from a lead can be Andreev reflected as a hole to the same lead, but will {\it never} appear in the other lead. In other words, no electron teleportation happens there. Indeed, Bolech and Demler has shown\cite{demler} that: (a) the conductance of each lead is $2e^2/h$ at zero bias and zero temperature, which is a sign of charge $2e$ transfer by Andreev reflection; (b) the two leads have independent currents without any correlation (no teleportation). The same results were also obtained using scattering approach within Bogoliubov-de Gennes formalism\cite{bk1}. In the two-terminal device we studied (Fig.2), charge $2e$ transfer is suppressed by charging energy and the conductance is at most $e^2/h$ because of single electron tunneling.


\begin{figure}
\centering
\includegraphics[width=3.5in]{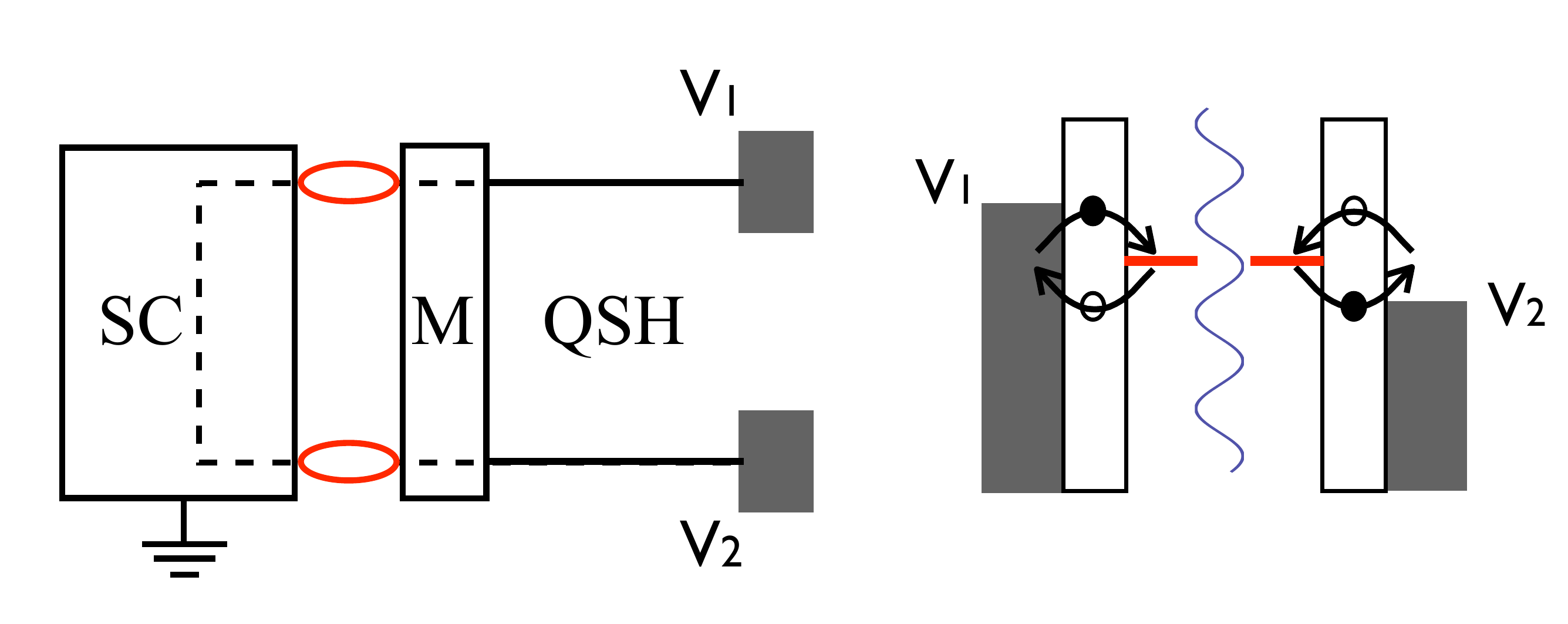}
\caption{Compared with Fig.2, the superconductor here is grounded without charging energy. Charge transfer between the lead and the superconductor is conducted by {\it local} Andreev reflection that transfers charge 2$e$. }
\end{figure}
 
 Finally, we discuss the feasibility of experiments using the quantum spin Hall insulator HgTe quantum well. A good candidate for the superconductor in our proposed setup is indium with $T_c=3.4$K, which is currently used as electrodes to contact HgTe\cite{HgTe-In}. The relevant parameters for these materials have been estimated\cite{fk3, bk1}.
Assuming a proximity-induced gap $\Delta \sim 0.1$meV, the penetration length of Majorana bound states is about $3{\mu}$m. 
So if the top and bottom Majorana bound states in Fig.2 are $30{\mu}$m apart, direct tunneling between them are negligible. 
If charging energy of the superconductor can be optimized to be comparable to $\Delta$, the resonant-tunneling model for electron teleportation described above is valid below 1K. The level broadening  $\Gamma$ from coupling to leads sets the temperature scale for detecting the phase-coherence.   

 
In summary, we reveal a striking non-local electron transport phenomena through Majorana bound states in a finite-sized superconductor with charging energy. 
Most interestingly, the transmission phase shift detects the state of a qbit made of two spatially separated Majorana bound states. In a future work\cite{future}, we will propose a generalized scheme to electrically detect the internal states of multiple Majorana bound states, with an emphasis on implementing topological quantum computation. 

We thank L. Glazman, F. D. M. Haldane, C. L. Kane, I. Neder, M. Rudner and especially B. I. Halperin for helpful discussions. 
We are grateful to A. Kitaev for an insightful conversation and B. I. Halperin for very valuable comments on the manuscript.
This work was supported by the Harvard Society of Fellows. 


\begin{thebibliography}{10}
\bibitem{gauge} We have chosen the gauge ${\vec A}=0$ everywhere in the superconductor.
\end{thebibliography}
\end{document}